\documentclass[aps,pra,twocolumn,groupedaddress]{revtex4-1}
\usepackage{graphicx,xcolor}
\usepackage{gensymb}
\usepackage{multirow}

\begin{document}

\title{First-principles insights into ultrashort laser spectroscopy of molecular nitrogen}

\author{Mohammad Reza Jangrouei}

\author{S. Javad Hashemifar}
\email{hashemifar@cc.iut.ac.ir}

\affiliation{Department of Physics, Isfahan University of Technology, 
             84156-83111 Isfahan, Iran}

\date{\today}
\newcommand{\etal}{{\em et al}}

\begin{abstract}
In this research, we employ accurate time-dependent density functional calculations for 
ultrashort laser spectroscopy of nitrogen molecule. 
Laser pulses with different frequencies, intensities, and durations are applied 
to the molecule and the resulting photoelectron spectra are analyzed. 
It is argued that relative orientation of the molecule in the laser pulse 
significantly influence the orbital character of the emitted photoelectrons. 
Moreover, the duration of the laser pulse is also found to be very effective in controlling 
the orbital resolution and intensity of photoelectrons. 
Angular resolved distribution of photoelectrons are computed at different 
pulse frequencies and recording times. 
By exponential growth of the laser pulse intensity, the theoretical threshold of 
two photons absorption in nitrogen molecule is determined.
\end{abstract}

\pacs{}
\keywords{Time dependent density functional theory, Ultrashort laser spectroscopy, 
          Nitrogen molecule, Photoelectron spectra, Two photon absorption}

\maketitle

\section{Introduction}

The recent progress in the field of ultra-short laser pulses has provided 
novel opportunities to capture fast dynamics of atoms and electrons in chemical reactions 
and photo ionization phenomena \cite{zewail2000,hentschel2001,krausz2009,lepine2014}. 
The studies of Ahmed Hassan Zewail, on the dynamic of chemical reactions 
by using femtosecond spectroscopy, 
awarded him the Nobel Prize of chemistry in 1999 \cite{zewail2000}. 
In 2001, Ferenc Krausz succeeded to generate attosecond laser pulses \cite{hentschel2001}, 
which provide invaluable abilities to investigate and even control electron dynamics 
in photo ionization phenomena \cite{pazourek2015}. 
Haessler and coworkers used a train of attosecond laser pulses in presence of 
a weak infrared field to ionize nitrogen molecule \cite{haessler2009}.
They identified two ionization channels in the system correspond to 
the ground state and excited state of the ionized molecule. Kelkensberg and others applied 
the same method to ionize hydrogen molecule and observe changes in charge distribution of 
the system on attosecond time scales \cite{kelkensberg2011}. 
Siu \etal. found that the time delay between attosecond pulse train and 
a corresponding infrared field may be used to control the dissociative ionization 
of oxygen molecule \cite{siu2011}. 
Penka and others applied time dependent density functional theory in 
the nonlinear nonperturbative regime to investigate laser induced photo ionization 
in CO and H$_2$CO molecules \cite{penka2014}. 
They found that the interplay between the ionization potential, 
the orbital shape, and the laser polarization axis significantly 
influence the ionization process.

In the present work, we employ time-dependent ab-initio calculations to study 
photo ionization of N$_2$ molecule under irradiation of short laser pulses. 
The effects of frequency and intensity of the pulse on the polarization will be investigated.

\section{Computational Method}

Our calculations have been performed in the framework of time dependent Kohn-Sham (TDKS)
density functional theory \cite{ullrich2011}, 
which provides a proper single-particle description of many-body
systems in the presence of time dependent external potentials (e.g. an electromagnetic pulse)
Adiabatic local density approximation (ALDA) is adapted for description of the time dependent
exchange correlation functional in this approach. It is already argued to be 
the proper functional for description of atomic clusters under 
intense electromagnetic fields \cite{fratalocchi2011}. 
We used the Octopus package to solve the TDKS equations by employing 
the norm-conserving pseudo potential technique \cite{castro2006}. 
The KS orbitals are expanded on a real space grid defined inside geometrical boxes
around atoms or around whole system. Two general approaches are implemented in this package for
solving the TDKS equations: linear response and explicit real time propagation methods. In the
linear response regime, which is used to address the effects of a weak uniform white
electromagnetic noise, the absorption spectra and the character of electronic excitations 
of the system are determined. 
While in the presence of strong laser pulses, explicit propagation of KS
orbitals in real time domain is considered.

In order to calculate the emitted photoelectron spectra of a sample after strong laser irradiation,
a detector region is defined around the system and then the Wigner quasi-probability distribution
function in the phase space:

$$\omega({\bf R},{\bf p},t)=\int{\frac{d{\bf s}}{2\pi^2}e^{i{\bf p}\cdot{\bf s}}
  \rho({\bf R}+\frac{{\bf s}}{2},{\bf R}-\frac{{\bf s}}{2},t)}$$

is used to integrate the photoelectrons in the detector region. 
In the above equation, $\rho({\bf r},{\bf r'},t)$ is a two body density matrix 
and $R$ and $s$ are the center of mass and relative coordinates. 
The momentum resolved photoelectron spectrum is then given by:

$$P({\bf p})=\lim_{t\rightarrow\infty}{\int{d{\bf R}\,\omega({\bf R},{\bf p},t)}}$$

In this equation the integral is calculated in the detector region after 
a sufficiently long time to ensure contribution of all photoelectrons. 
In the Kohn-Sham approach, the two body density matrix is defined by the following
sum over occupied states:

$$\rho_{\rm KS}({\bf r},{\bf r'},t)=\sum_{i}^{occ.}{\psi_i({\bf r},t)\psi_i({\bf r'},t)}$$

This procedure needs calculation area of hundreds Angstrom to give reliable
photoelectron spectra. In order to reduce the required calculation area, 
a mask region is defined before the detector region \cite{de2012}.

\begin{table}
\caption{\label{states}
  Obtained molecular orbitals of the nitrogen molecule. 
  The Occupied orbitals are highlighted.
}
\begin{ruledtabular}
\begin{tabular}{|clcc|ccc|}
state &     energy (eV)        &&& state & energy (eV) &  \\
\hline                           
\bf 1 &\bf -28.40~($\sigma  $) &&&  11   &   0.53 &  \\ 
\bf 2 &\bf -13.38~($\sigma^*$) &&&  12   &   1.05 &  \\
\bf 3 &\bf -11.96~($\pi     $) &&&  13   &   1.12 &  \\
\bf 4 &\bf -11.96~($\pi     $) &&&  14   &   1.12 &  \\
\bf 5 &\bf -10.45~($\sigma  $) &&&  15   &   1.13 &  \\
    6 &   ~~-2.32~($\pi^*$)    &&&  16   &   1.13 &  \\
    7 &   ~~-2.32~($\pi^*$)    &&&  17   &   1.21 &  \\
    8 &   ~~~0.03~($\sigma^*$) &&&  18   &   1.65 &  \\
    9 &   ~~~0.52              &&&  19   &   1.65 &  \\
   10 &   ~~~0.53              &&&  20   &   1.66 &  \\
              
\end{tabular}
\end{ruledtabular}
\end{table}

\section{Results and Discussions}

\begin{figure}
\includegraphics[scale=0.85]{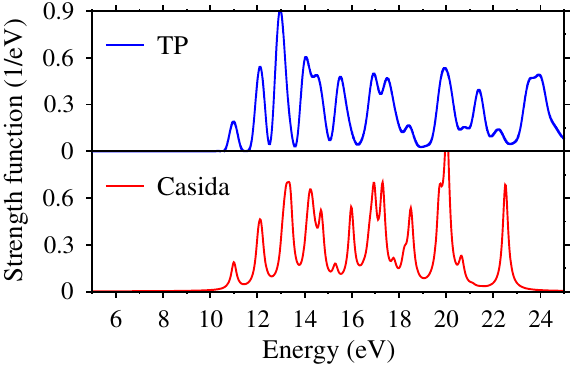}
\caption{\label{spectrum}
  Calculated absorption spectra of N2 by using the Casida linear response 
  and real time propagation (TP) methods.
}
\end{figure}

First, we performed some static DFT calculations to identify the equilibrium properties 
of N$_2$ molecule. The equilibrium bond length and binding energy of nitrogen was found 
to be about 1.09\,\AA\ and 8.89\,eV, respectively, 
which agree with the measured data (1.1\,\AA\ and 9.79\,eV) \cite{lide2006}.
The energy gap between the highest occupied molecular orbital (HOMO) and 
the lowest unoccupied molecular orbital (LUMO) was determined to 
be about 8.2\,eV (table~\ref{states}). 
Comparing this parameter with the experimental energy gap of N$_2$ is questionable, 
because of the frozen character of orbitals in the static DFT calculations, 
while in practice; electron excitation has non-trivial influences on orbital energy levels.
This problem is well resolved in time dependent DFT, where orbitals are allowed 
to relax during electronic excitations. 
The obtained absorption spectrums of N$_2$ by using TDDFT within the Casida
linear response and real time propagation approaches are presented in Fig.~\ref{spectrum}. 
The agreement between these two spectra is acceptable, especially in lower energies. 
In higher energies, the accuracy of the linear response approach decreases and 
hence the real time propagation approach is more reliable.

The obtained energy gap within both Casida and real time approaches is about 11.0\,eV. 
In order to compare the obtained absorption spectra with experiment, 
we note the complex absorption spectrum of nitrogen molecule \cite{vieitez2008}, 
which includes weak dipole-forbidden transitions from 6 to 12.4\,eV 
and strong dipole-allowed transitions from 12.4 to 18.8\,eV \cite{hudson1971}.
The first 20 electronic excitations, identified within the Casida approach, 
are listed in table~\ref{excitations}. Our results confirm that the
peaks below 11\,eV have negligible dipole moment and very weak strength while strong
dipole allowed peaks occur above this threshold. Moreover, we may conclude that the observed
experimental transitions below 11\,eV are likely non-electronic excitation transition 
(rotational or vibrational transitions). 
The ionization energy of nitrogen molecule, the difference of the minimized
energy of the neutral and ionized molecule, was found to be 16.02\,eV,
which compares well with the measured value of 15.80\,eV\ \cite{radzig1986}.

\begin{table*}
\caption{\label{excitations}
  Identified characteristic parameters of the first 20 electronic excitations of N$_2$,
  within the Casida linear response approach, energy (eV): excitation energy,
  dipole (\AA): Cartesian components of the excitation dipole moment,
  strength (1/eV): excitation strength,
  character: major involved transitions between molecular orbitals, 
  the corresponding probability domains are written in the parenthesis and 
  molecular orbital numbers are consistent with table \ref{states}.
  The dipole allowed transitions are highlighted.
}
\newcommand{\ra}{{$\rightarrow$}}
\begin{ruledtabular}
\begin{tabular}{ccccccccl}
      & energy  &&  \multicolumn{3}{c}{dipole}     &&  strength  &    character \\
\hline                                                       
     1&     9.19&&  1.74E-09 & 5.48E-06 & 4.58E-06 &&    4.10E-11&    5\ra6~(0.74),  5\ra7~(-0.66) \\
     2&     9.19&&  1.27E-09 & 4.58E-06 & 5.48E-06 &&    4.10E-11&    5\ra6~(0.66),  5\ra7~(0.74)  \\
     3&     9.69&&  5.67E-12 & 5.71E-11 & 6.62E-11 &&    6.51E-21&    4\ra6~(0.71),  3\ra7~(0.71)  \\
     4&    10.23&&  7.27E-09 & 3.22E-10 & 1.08E-09 &&    4.85E-17&    3\ra6~(0.68),  4\ra6~(-0.19), 3\ra7~(0.19), 4\ra7~(0.68) \\
     5&    10.24&&  2.16E-09 & 1.09E-09 & 3.09E-10 &&    5.33E-18&    3\ra6~(-0.19), 4\ra6~(-0.68), 3\ra7~(0.68), 4\ra7~(-0.19)\\
     6&    10.65&&  8.47E-07 & 3.17E-10 & 8.69E-10 &&    6.68E-13&    5\ra8~(0.99)   \\
                  
\bf  7&\bf 11.00&&\bf 2.05E-01&    1.39E-09&    5.84E-09&&\bf 4.05E-02&\bf 5\ra9~(1.00)   \\
\bf  8&\bf 11.00&&    7.55E-09&\bf 2.86E-02&\bf 1.29E-01&&\bf 1.68E-02&\bf 5\ra10~(0.12), 5\ra11~(0.99)  \\
\bf  9&\bf 11.00&&    3.81E-09&\bf 1.29E-01&\bf 2.86E-02&&\bf 1.68E-02&\bf 5\ra10~(0.99), 5\ra11~(-0.12) \\
                  
    10&    11.56&&  9.73E-07 & 1.36E-10 & 5.08E-10 &&    9.57E-13&    5\ra12~(0.97), 5\ra17~(0.25)  \\
    11&    11.61&&  3.08E-10 & 4.52E-08 & 8.81E-07 &&    7.91E-13&    5\ra14~(1.00)  \\
    12&    11.61&&  5.60E-11 & 8.81E-07 & 4.52E-08 &&    7.91E-13&    5\ra13~(1.00)  \\
    13&    11.62&&  6.49E-10 & 3.93E-11 & 4.41E-11 &&    4.31E-19&    5\ra15~(1.00)  \\
    14&    11.62&&  3.20E-10 & 2.33E-11 & 1.69E-11 &&    1.05E-19&    5\ra16~(1.00)  \\
    15&    11.83&&  3.38E-07 & 1.65E-10 & 1.19E-09 &&    1.19E-13&    5\ra8~(-0.11), 5\ra12~(-0.24), 5\ra17~(0.96) \\
                  
\bf 16&\bf 12.04&&    3.90E-11&\bf 2.02E-01&\bf 2.51E-02&&\bf 4.38E-02&\bf 4\ra8~(0.99),  5\ra18~(-0.11) \\
\bf 17&\bf 12.04&&    1.48E-09&\bf 2.512-02&\bf 2.02E-01&&\bf 4.38E-02&\bf 3\ra8~(0.99),  5\ra19~(-0.11) \\
\bf 18&\bf 12.13&&    3.74E-09&\bf 1.02E-02&\bf 1.20E-01&&\bf 1.53E-02&\bf 3\ra8~(0.11),  5\ra19~(0.99), 5\ra25~(0.11)  \\
\bf 19&\bf 12.13&&    3.41E-09&\bf 1.20E-01&\bf 1.02E-02&&\bf 1.53E-02&\bf 4\ra8~(0.11),  5\ra18~(0.99), 5\ra24~(-0.11) \\
\bf 20&\bf 12.14&&\bf 2.51E-01&    2.44E-09&    5.64E-10&&\bf 6.67E-02&\bf 5\ra20~(0.99), 5\ra23~(-0.16) \\
\end{tabular}
\end{ruledtabular}
\end{table*}

\begin{figure*}
\includegraphics{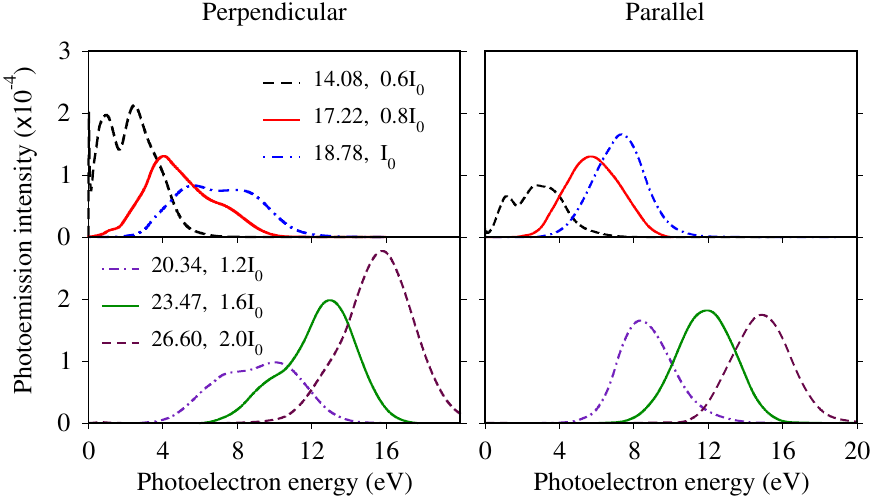}
\caption{\label{pes}
  Calculated photoelectron spectra of N$_2$ at the perpendicular and parallel
  geometries at the pulse intensity of $10^{14}$\,W/cm$^2$ and 
  at six different laser pulse frequencies from 14.08 to 26.60\,eV. 
}
\end{figure*}

For calculations of photoelectron spectra, we used spherical boxes around the molecule with an
optimum internal radius of 12\,\AA\ for region A, an external radius of 22\,\AA\ for 
detector region and a sine-mask function. 
The optimum grid spacing in the atomic spheres was found to be 0.18\,\AA\ while 
the time step for real time evolution was set to 1\,m$\hbar$/eV ($\sim$0.66\,as). 
We used Gaussian envelope laser pulses with a length of 4\,$\hbar/$eV (2.63\,fs). 
The frequency of the extreme ultraviolet (xuv) laser pulses was set 
to some specific odd (9-17$^{\rm th}$) multiples of a fundamental frequency of 1.565\,eV.
These odd harmonics have been already produced by propagating intense laser pulses in a
gas jet and then used for photo ionization of nitrogen molecule \cite{haessler2009}. 
The 12$^{\rm th}$ multiple was also considered for more accurate inspection.

\begin{figure*}
\includegraphics{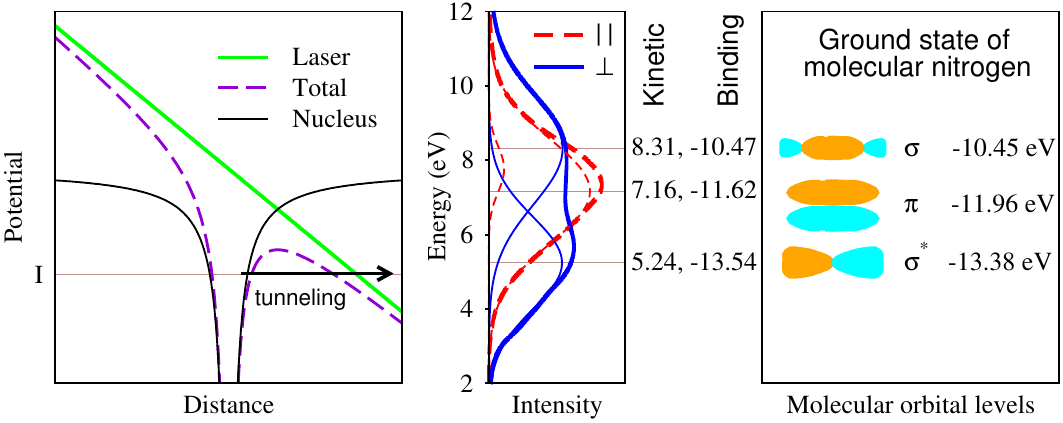}
\caption{\label{tunnel}
  left:   schematic representation of photo ionization via electron tunneling in 
          a strong laser field. The letter I stand for ionization potential. 
  Middle: calculated photoelectron spectra in the perpendicular (solid line) 
          and parallel (dashed line) geometries at pulse frequency of 18.78\,eV 
          and pulse intensity of $10^{14}$\,W/cm$^2$. 
          The spectrum of perpendicular geometry is 50\% enlarged to be clearer. 
          The photoelectrons kinetic energy is identified by deconvolution of the spectra 
          in two Gaussian functions. The kinetic energy of the photoelectrons is subtracted 
          from the laser pulse energy to determine the binding energy of photoelectrons. 
  Right:  the highest occupied energy levels of molecular nitrogen in the ground state. 
}
\end{figure*}

The calculated photoelectron spectra at the desired pulse frequencies and two different
geometries are presented in Fig.~\ref{pes}. 
In these geometries, the molecule is either parallel or perpendicular to 
the direction of the laser pulse propagation. 
First, we focus on the perpendicular geometry. 
It is seen that at the lowest pulse frequency (14.08\,eV), two peaks are appeared 
in the spectra, indicating emission of two different kinds of photoelectrons 
from the system. 
We will argue that these peaks are likely attributed to the two sigma molecular orbitals 
of the nitrogen molecule.
Taking into account the calculated ionization energy of N$_2$ (16.02\,eV),
it seems that a laser pulses with frequency of 14.08\,eV should not be able 
to create any photoelectrons. 
Therefore, the observed very weak ionization is either due to the multi-photon absorption 
or electron tunneling in strong laser field. 
The low kinetic energy of photoelectrons ($\sim 2$\,eV) rules out the multi-photon
absorption mechanism. The suppressed electrostatic potential of a nucleus in 
a strong laser field is depicted in Fig.~\ref{tunnel}. 
It is clearly seen that the laser field may reduce the electrostatic confinement
barrier of the nucleus and hence enhances electron tunneling. 
Therefore tunneling ionization may happen at energies lower than 
the normal ionization energy. The very low intensity of the photoelectron spectra 
(Fig.~\ref{pes}) provides further evidence for the tunneling mechanism of ionization 
in our system.

At higher laser frequencies, compared with 14.08\,eV, photoelectrons gain 
more kinetic energy and hence the spectra shifts to higher energies. 
We observe that, in the perpendicular geometry, the two peaks occur at all frequencies, 
although in some cases one of the peaks is weaker and hence appears 
as a shoulder in the spectra. 
On the other hand, in the parallel geometry, except for the first frequency, 
in all other frequencies only one peak is visible in the spectra. 
In order to understand this feature, we focus on the laser frequency of 18.78\,eV
and identify the peak positions in the perpendicular and parallel geometries (Fig.~\ref{tunnel}). 
In this regard, we have tried to deconvolute the photoelectron spectra in two
Gaussian functions. The center of the Gaussian peaks may be assigned to 
the kinetic energy of the photoelectrons. 
It is seen that (Fig.~\ref{tunnel}) the single peak of the parallel geometry 
is decomposed into a major peak around 7.16\,eV and a minor peak at 7.80\,eV. 
On the other hand, in the perpendicular geometry two peaks at 5.24 and 8.31\,eV 
are contributing to the spectra. 
The major peak of the parallel geometry is located between the two peaks of 
the perpendicular geometry. By subtracting the kinetic energy of photoelectrons 
from the laser pulse energy, we may determine the binding energy of the emitted electrons. 
Then, we consider the three highest occupied molecular orbitals of
N$_2$, one $\pi$ molecular orbital located between two $\sigma$ orbitals 
(table~\ref{states}, Fig.~\ref{tunnel}). 
These orbital levels exhibit very good consistency with the obtained binding energy 
of the photoelectrons. 
This consistency enables us for a brief anatomy of the calculated photoelectron spectra.
The $\sigma$ orbitals are mainly distributed along the molecular axis, 
while in the case of $\pi$ orbital, out of axis distribution may also play a significant role. 
In the perpendicular geometry, the pulse electric field is along the molecular axis 
and hence mainly $\sigma$ and $\sigma^*$ photoelectrons are emitted from the system. 
In the parallel geometry, the pulse field is perpendicular to axis and hence mainly ionizes
the $\pi$ orbital of the molecule, with a minor contribution from the $\sigma$ orbital. 
These arguments presumably explain occurrence of one (two) peaks in the photoelectron spectra 
of the molecular nitrogen in the parallel (perpendicular) geometries.

The intensity of the photoelectron spectra exhibits different trends in the parallel and
perpendicular geometries (Fig.~\ref{pes}). 
In the parallel geometry, we observe that by increasing the pulse frequency, 
the photoelectron spectra intensity increases smoothly. 
But in the perpendicular geometry, a more complicated trend is seen. 
The photoelectron intensity decreases in the frequency range of 14.08 to 18.78\,eV 
and then increases from 20.34 to 26.60\,eV. 
In fact, theoretical description of strong field ionization has already shown that 
tunneling rate is a complicated function of the laser pulse frequency 
and intensity \cite{popruzhenko2008}.

\begin{figure}
\includegraphics[scale=0.9]{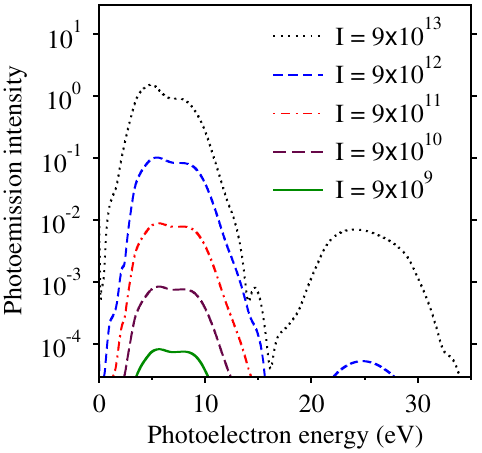}
\caption{\label{tpa}
  Calculated photoelectron spectra of N$_2$ at five different laser pulse
  intensities from $10^{14}$ to $10^{18}$ W/cm$^2$. 
  The pulse frequency was set to 18.78\,eV.
}
\end{figure}

In order to investigate feasibility of multi photon absorption in molecular nitrogen, 
we considered laser pulses with frequency of 18.78\,eV and five different intensities 
from $10^{13}$ to $10^{18}$\,W/cm$^2$. 
The obtained results are presented in Fig.~\ref{tpa}. 
For the intensities lower than $10^{17}$\,W/cm$^2$, the general feature of the spectra 
does not change by increasing the pulse intensity. 
The spectra have two peaks and the photoelectron intensity is linearly scaled 
by the pulse intensity. 
At the pulse intensity of $10^{17}$\,W/cm$^2$, a third peak appears in the spectra 
with a kinetic energy of about 25\,eV.
Adding this value to the binding energy of $\sigma$ orbital (10.47\,eV) gives 
a minimum required energy of about 35.5\,eV for emission of the corresponding photoelectrons, 
which is almost twice the energy of a single laser photon (18.78\,eV). 
Hence, we conclude that at this high intensity two photons absorption happens in the system. 
At the pulse intensity of $10^{18}$\,W/cm$^2$, the intensity of the third peak 
increases about two order of magnitudes compared with the previous one. 
Hence, the two photons absorption intensity scales with the square of 
the laser pulse intensity, in well agreement with theoretical description of 
this nonlinear optical phenomenon \cite{shen1984}. 
In the presence of high intensity laser pulses, the pondermotive energy 
may also influence the results.
In the case of long pulses the pondermotive energy has a well
defined constant value $\epsilon^2/4\omega^2$ where
$\epsilon$ is the electric field amplitude \cite{de2012}.
However, in the case of intense ultrashort pulses it is argued that
the pondermotive energy is not constant and calculation of this
parameter is not straightforward \cite{della2016}.
Moreover, it is discussed that in this situation the Stark shift
may substantially compensate the pondermotive shift \cite{fushitani2016}.
Hence, we ignore consideration of non-constant pondermotive energy
in the current work.

\begin{figure}
\includegraphics[trim=30 120 0 30 , scale=0.4]{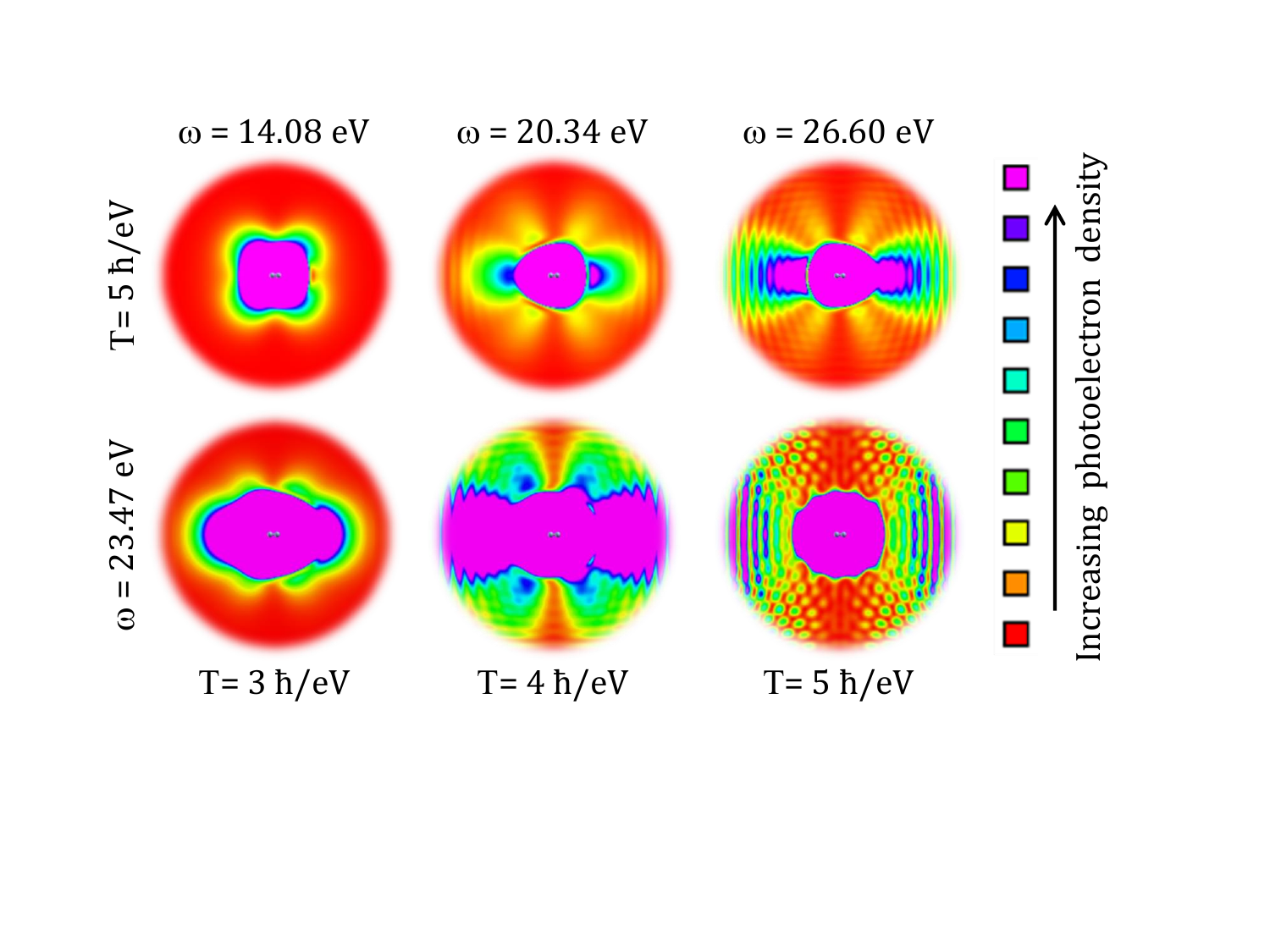}
\caption{\label{angular}
  Effects of pulse frequency and recording time on the angular distribution
  of photoelectrons emitted from nitrogen molecule. 
  The top row shows the distributions at three different pulse frequencies 
  at the recording time of 5\,$\hbar$/eV. after pulse irradiation. 
  The bottom row shows the photoelectron distributions at a single pulse frequency 
  and at three different recording times. 
  The laser pulse intensities are the same as Fig.~\ref{pes}.
}
\end{figure}

The angular distribution of the photoelectrons resulted from the laser pulse 
frequencies of 14.08, 20.34 and 26.60\,eV are presented in Fig.~\ref{angular}. 
Obviously, increasing frequency of the incident pulse enhances the kinetic energy 
of the photoelectrons and hence speeds up its propagation away from the molecule. 
The laser pulse is propagating perpendicular to the molecule and the electric field 
polarization is parallel to the molecule. 
Hence, we observe that photoelectrons are propagating in the molecule direction. 
We have also investigated time evolution of the photoelectron distribution. 
At the laser pulse frequency of 23.47\,eV, the distribution is recorded after three
propagation times (3, 4, and 5\,$\hbar$/eV). 
We observe that before arriving to the detector wall, the photoelectron density 
has a smooth distribution, however after incident to the detector wall, 
many fluctuations appear in the distribution density.

\begin{figure}
\includegraphics[scale=0.9]{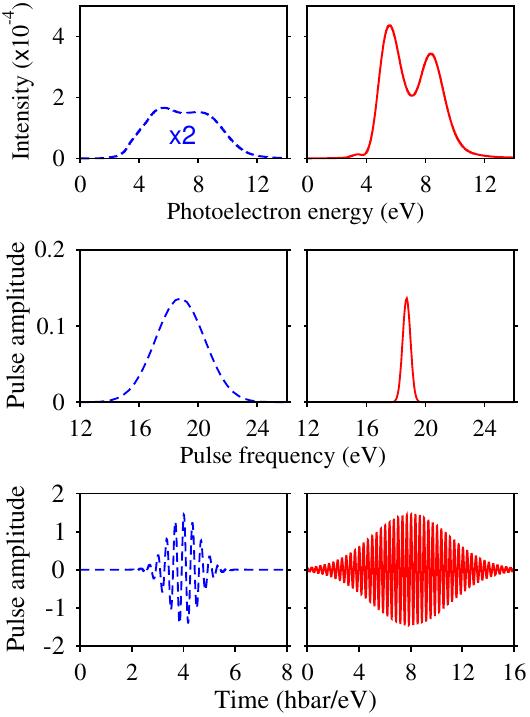}
\caption{\label{duration}
  Effects of pulse duration on the photoelectrons spectra of molecular nitrogen. 
  Two pulse lengths of 4 and 16\,$\hbar$/eV, their Fourier transform, 
  and the resulting photoelectron spectra have been compared together. 
  The frequency of the pulses is 18.78\,eV. 
  The photoelectron spectrum of the shorter pulse is enlarged twice to be clearer.
}
\end{figure}

Throughout this project, we have mainly used ultrashort laser pulses with 
a duration of 4\,$\hbar$/eV ($\sim 2.6$\,fs) to mimic the real experimental situations. 
In order to see the effect of pulse duration on the photoelectron spectra, 
a laser pulse with a four times longer duration ($\sim 10.5$\,fs) were considered in
our study (Fig.~\ref{duration}). 
In general, the energy uncertainty principle, $\Delta E\,\Delta t\leq\hbar$,
applies that increasing the pulse duration should decrease the energy tolerance. 
As it is obvious in the figure, the Fourier transform of the longer pulse 
has much narrower dispersion around the central frequency of 18.78\,V. 
Therefore, the resulting photoelectron spectrum has more resolution in terms of 
orbital character of the emitted photoelectrons. 
We observe that the two peaks of the corresponding spectrum are significantly sharper, 
compared with the photoelectron spectrum of the shorter pulse. 
Moreover, frequency tolerance of the pulse exhibits nontrivial influence on 
the intensity of the photoelectrons.
A laser pulse with sharper frequency distribution is clearly much more efficient 
for electron emission from the sample. 

\section{Conclusions}

Real time propagation of the single particle Kohn-Sham orbitals within 
adiabatic local density approximation was applied to study photoelectron spectra 
of nitrogen molecule in short laser pulses. 
It was argued that when direction of the pulse propagation is perpendicular to the molecule, 
$\sigma$ photoelectrons are mainly emitted from the system, 
while in the parallel geometry the highest occupied $\pi$ orbital is more ionized. 
It was seen that longer laser pulses, with lower frequency dispersion, 
are more efficient for creation of well orbital resolved photoelectrons. 
Angular resolved distributions were plotted to observe real space propagation of
photoelectrons at different pulse frequencies and propagation time. 
It was argued that at $10^{17}$\,W/cm$^2$ pulse intensities and higher, 
some new photoelectrons with much higher kinetic energy are emitted from the
molecule which indicate occurrence of two photons absorption phenomenon.

\begin{acknowledgments}
This work was supported by the Vice Chancellor of Isfahan University of Technology (IUT) 
in research affairs. 
SJH acknowledges the Abdus Salam International Center for Theoretical Physics
(ICTP) for supporting his summer (2016) visit to ICTP.
\end{acknowledgments}

\end{document}